\documentclass[12pt]{article}
\usepackage{fancyhdr}

\usepackage[normalem]{ulem}
\usepackage{mathrsfs}
\usepackage[T1]{fontenc}
\usepackage{mathpazo}
\usepackage{setspace}
\usepackage{amsfonts}
\usepackage{amssymb}
\usepackage{amsmath}
\usepackage{epsfig}
\usepackage{latexsym}
\usepackage{color}
\usepackage{graphicx}
\usepackage{nicefrac}
\usepackage[latin1]{inputenc}
\usepackage{slashed}
\usepackage{multirow}
\usepackage{datetime}
\usepackage{hyperref}
\usepackage{comment}
\usepackage{cite}
\usepackage{soul}

\def\hybrid{\topmargin -20pt    \oddsidemargin 0pt
        \headheight 0pt \headsep 0pt
        \textwidth 6.25in       % A4 paper
        \textheight 9.25in       % A4 paper
        \marginparwidth .875in
        \parskip 5pt plus 1pt   \jot = 1.5ex}

\hybrid

\def\baselinestretch{1.2}

\catcode`\@=11

\def\marginnote#1{}
\newcount\hour
\newcount\minute
\newtoks\amorpm
\hour=\time\divide\hour by60
\minute=\time{\multiply\hour by60 \global\advance\minute by-\hour}
\edef\standardtime{{\ifnum\hour<12 \global\amorpm={am}%
        \else\global\amorpm={pm}\advance\hour by-12 \fi
        \ifnum\hour=0 \hour=12 \fi
        \number\hour:\ifnum\minute<10 0\fi\number\minute\the\amorpm}}
\edef\militarytime{\number\hour:\ifnum\minute<10 0\fi\number\minute}
\def\draftlabel#1{{\@bsphack\if@filesw {\let\thepage\relax
   \xdef\@gtempa{\write\@auxout{\string
      \newlabel{#1}{{\@currentlabel}{\thepage}}}}}\@gtempa
   \if@nobreak \ifvmode\nobreak\fi\fi\fi\@esphack}
        \gdef\@eqnlabel{#1}}
\def\@eqnlabel{}
\def\@vacuum{}
\def\draftmarginnote#1{\marginpar{\raggedright\scriptsize\tt#1}}

\def\draft{\oddsidemargin -.5truein
        \def\@oddfoot{\sl preliminary draft \hfil\mathrm{d}s
        \rm\thepage\hfil\sl\today\quad\militarytime}
        \let\@evenfoot\@oddfoot \overfullrule 3pt
        \let\label=\draftlabel
        \let\marginnote=\draftmarginnote
   \def\@eqnnum{(\theequation)\rlap{\kern\marginparsep\tt\@eqnlabel}%
\global\let\@eqnlabel\@vacuum}  }

\def\preprint{\twocolumn\sloppy\flushbottom\parindent 2em
        \leftmargini 2em\leftmarginv .5em\leftmarginvi .5em
        \oddsidemargin -.5in    \evensidemargin -.5in
        \columnsep .4in \footheight 0pt
        \textwidth 10.in        \topmargin  -.4in
        \headheight 12pt \topskip .4in
        \textheight 6.9in \footskip 0pt
        \def\@oddhead{\thepage\hfil\addtocounter{page}{1}\thepage}
        \let\@evenhead\@oddhead \def\@oddfoot{} \def\@evenfoot{} }

\def\numberbysection{\@addtoreset{equation}{section}
        \def\theequation{\thesection.\arabic{equation}}}

\def\underline#1{\relax\ifmmode\@@underline#1\else
        $\@@underline{\hbox{#1}}$\relax\fi}

\def\titlepage{\@restonecolfalse\if@twocolumn\@restonecoltrue\onecolumn
     \else \newpage \fi \thispagestyle{empty}\c@page\z@
        \def\thefootnote{\fnsymbol{footnote}} }

\def\endtitlepage{\if@restonecol\twocolumn \else \newpage \fi
        \def\thefootnote{\arabic{footnote}}
        \setcounter{footnote}{0}}  %\c@footnote\z@ }

\catcode`@=12
\relax

\def\figcap{\section*{Figure Captions\markboth
        {FIGURECAPTIONS}{FIGURECAPTIONS}}\list
        {Figure \arabic{enumi}:\hfill}{\settowidth\labelwidth{Figure
999:}
        \leftmargin\labelwidth
        \advance\leftmargin\labelsep\usecounter{enumi}}}
 \relax
\def\tablecap{\section*{Table Captions\markboth
        {TABLECAPTIONS}{TABLECAPTIONS}}\list
        {Table \arabic{enumi}:\hfill}{\settowidth\labelwidth{Table
999:}
        \leftmargin\labelwidth
        \advance\leftmargin\labelsep\usecounter{enumi}}}
 \relax
\def\reflist{\section*{References\markboth
        {REFLIST}{REFLIST}}\list
        {[\arabic{enumi}]\hfill}{\settowidth\labelwidth{[999]}
        \leftmargin\labelwidth
        \advance\leftmargin\labelsep\usecounter{enumi}}}
 \relax
\makeatletter
\newcounter{pubctr}
\def\publist{\@ifnextchar[{\@publist}{\@@publist}}
\def\@publist[#1]{\list
        {[\arabic{pubctr}]\hfill}{\settowidth\labelwidth{[999]}
        \leftmargin\labelwidth
        \advance\leftmargin\labelsep
        \@nmbrlisttrue\def\@listctr{pubctr}
        \setcounter{pubctr}{#1}\addtocounter{pubctr}{-1}}}
\def\@@publist{\list
        {[\arabic{pubctr}]\hfill}{\settowidth\labelwidth{[999]}
        \leftmargin\labelwidth
        \advance\leftmargin\labelsep
        \@nmbrlisttrue\def\@listctr{pubctr}}}
 \relax
\makeatother
\newskip\humongous \humongous=0pt plus 1000pt minus 1000pt

\newif\ifdtup

\relax

\def\be{\begin{equation}}
\def\ee{\end{equation}}
\def\ba{\begin{eqnarray}}
\def\ea{\end{eqnarray}}

\def\del{\partial}

\def\k{\kappa}

\def\a{\alpha}

\def\b{\beta}

\def\g{\gamma}

\def\d{\delta}
\def\D{\Delta}

\def\m{\mu}

\def\om{\omega}

\def\l{\lambda}

\def\s{\sigma}
\def\S{\Sigma}

\def\cN{{\cal N}}

\def\cR{{\cal R}}

\def\no{\noindent}

\def\qq{\qquad}

\def\IR{\relax{\rm I\kern-.18em R}}

\def \ha {{1\over 2}}

\def \ov {\over}

\def\IR{\relax{\rm I\kern-.18em R}}
\def\IL{\relax{\rm I\kern-.18em L}}

\def\inv{^{\raise.15ex\hbox{${\scriptscriptstyle -}$}\kern-.05em 1}}

\def\cR{{\cal R}}

\def\R{{\cal{R}}}

\DeclareSymbolFont{bbold}{U}{bbold}{m}{n}
\DeclareSymbolFontAlphabet{\mathbbold}{bbold}
\begin{document}

\renewcommand{\theequation}{\thesection.\arabic{equation}}
\csname @addtoreset\endcsname{equation}{section}

\newcommand{\beq}{\begin{equation}}
\newcommand{\eeq}[1]{\label{#1}\end{equation}}
\newcommand{\ber}{\begin{eqnarray}}
\newcommand{\eer}[1]{\label{#1}\end{eqnarray}}
\newcommand{\eqn}[1]{(\ref{#1})}
\begin{titlepage}
\begin{center}

\vskip  1in

%\today

{\large \bf  Generalised  integrable  $\lambda$- and $\eta$-deformations and their relation}

\vskip 0.4in

{\bf Konstantinos Sfetsos,}$^{1,2}$\ {\bf Konstantinos Siampos}$^{2}$ and {\bf Daniel C. Thompson}$^{3}$
\vskip 0.1in
{\em ${}^1$Department of Nuclear and Particle Physics,\\
Faculty of Physics, University of Athens,\\
Athens 15784, Greece}
\vskip 0.1in
{\em ${}^2$Albert Einstein Center for Fundamental Physics,\\
Institute for Theoretical Physics, Bern University,\\
Sidlerstrasse 5, CH3012 Bern, Switzerland}
\vskip 0.1in
{\em ${}^3$Theoretische Natuurkunde, Vrije Universiteit Brussel,\\
and The International Solvay Institutes\\
Pleinlaan 2, B-1050, Brussels, Belgium}

{\tt\footnotesize ksfetsos@phys.uoa.gr, siampos@itp.unibe.ch, daniel.thompson@vub.ac.be}

\end{center}
\vskip 0.4in
\centerline{\bf Abstract}
We construct two-parameter families of integrable $\lambda$-deformations of two-dimensional field theories.
These interpolate between a CFT (a WZW/gauged WZW model) and the non-Abelian T-dual of a principal chiral model on a
group/symmetric coset space.
In examples based on the  $SU(2)$ WZW model and the $SU(2)/U(1)$ exact coset CFT,
we show that these deformations are related to bi-Yang--Baxter generalisations of $\eta$-deformations via Poisson--Lie T-duality and analytic continuation.  We illustrate the quantum behaviour of our models under RG flow.  As a byproduct we demonstrate that the  bi-Yang--Baxter $\s$-model for a general group is one-loop renormalisable.

\no

\newpage

\noindent

\vskip .4in
\noindent
\end{titlepage}
\vfill
\eject

\def\baselinestretch{1.2}
\baselineskip 20 pt
\noindent

%%%%%%%%%%%%%%%

\tableofcontents
\eject

\setcounter{equation}{0}
\renewcommand{\theequation}{\thesection.\arabic{equation}}

\section{Introduction and motivation}

One of the most powerful tools available to the modern holographic practitioner is integrability.  Most famously,  the problem of determining the anomalous dimensions of single trace operators in the planar limit of ${\cal N}=4$ supersymmetric Yang--Mills gauge theory with gauge group $SU(N)$ can be mapped to the problem of determining eigenvalues of an integrable spin-chain Hamiltonian \cite{Minahan:2002ve}.   On the other side of the $AdS/CFT$ conjecture, the $AdS_5\times S^5$ string $\sigma$-model is, classically at least, integrable.  The reason for this is that the $\sigma$-model's target space is exceptionally symmetric; the world sheet theory takes the form of a {$\sigma$-model} on a semi-symmetric space $PSU(2,2|4)/SO(4,1)\times SO(5)$ \cite{Metsaev:1998it}. The two-dimensional $\sigma$-model admits a Lax pair formulation from which an infinite tower of conserved quantities can be deduced \cite{Bena:2003wd}.

Given this success, one would hope to find ways in which the AdS/CFT correspondence can be generalised from the $AdS_5\times S^5$ setting whilst still maintaining the properties of   integrability.  Two novel and related classes of two-dimensional $\sigma$-models, that we shall refer to as $\eta$- and $\lambda$-deformations, have recently been developed and provide a new perspective on this challenge.

The $\eta$-deformation of the $AdS_5\times S^5$ superstring proposed by Delduc, Magro and Vicedo \cite{Delduc:2013fga,Delduc:2013qra} is a generalisation of the Yang--Baxter (YB) deformations introduced by Klim\v{c}\'ik in \cite{Klimcik:2002zj}. A central r\^ole in the construction of such YB deformations is played by the antisymmetric $\R$-matrix; an endomorphism of a Lie-algebra $\frak{g}$ that obeys a modified YB (mYB) equation
\be
\label{mYB}
[\R A, \R B] - \R([\R A,B]+[A,\R B] ) = -c^2[A, B]  \ ,  \quad \forall  A,B \in \frak{g} \,,\quad c\in\mathbb{C}\,.
\ee
There are three distinct choices for the parameter $c$; $c^2>0$, $c^2 < 0$ and $c^2 = 0$ and the corresponding solutions of the mYB  are referred to as being, respectively,  on the real, complex and classical branch.\footnote{
\label{jsksls}
Contracting  \eqref{mYB1},
equivalent form of \eqref{mYB}, with $f_{abc}$ and using the Jacobi identity we easily find that
\begin{equation*}
 c_G\, c^2\, {\rm dim}G+\frac32\,||\xi||^2=0\,,\qquad ||\xi||^2=\delta_{ab}\,\xi_a\xi_b\,,\qquad \xi_a=f_{abc}\,\R_{bc}\,,
\end{equation*}
which has no solution for compact groups and $c^2=1$, referred to as the real branch.
}
   The complex branch, $c^2<0$, is the setting for the $\eta$-deformations. Using such an ${\cal R}$-matrix one can construct a one-parameter family of deformations of the  principal chiral model on a group $G$ which were shown in  \cite{Klimcik:2002zj,Klimcik:2008eq} to be integrable.  This approach was generalised, and integrability shown,    for symmetric cosets in \cite{Delduc:2013fga} and for semi-symmetric spaces in  \cite{Delduc:2013qra}.   These $\eta$-deformations are particularly interesting since although the corresponding target spaces only display an Abelian subset of the original $AdS_5\times S^5$ isometry group, it is thought that the full symmetries of the string $\sigma$-model are governed by a quantum-group with  a real quantum-group parameter $q = e^{f(\eta)}$ \cite{Delduc:2013fga}, where $f$ is a real function of $\eta$, and perturbative evidence for this has been given in \cite{Arutyunov:2013ega}.

The $\lambda$-deformation was introduced by one of the present authors in \cite{Sfetsos:2013wia} and can be realised as an integrable interpolation between an exact CFT (a WZW/gauged-WZW model) and the non-Abelian T-dual of the principal chiral model on a group/coset space.
{In the bosonic case this deformation} is constructed by applying a gauging procedure to the combination of a PCM on a group (coset) and a (gauged)-WZW model. The deformation parameter is given in terms of the radius of the PCM $\kappa^2$, and the WZW level $k$, by
\begin{equation}
\lambda = \frac{k}{k+ \kappa^2}  \ .
\end{equation}
{
For cosets, this construction was initiated in \cite{Sfetsos:2013wia} (where more emphasis was given to the cases corresponding to
group spaces), and performed more rigorously for
symmetric coset spaces in \cite{Hollowood:2014rla} and further generalised to semi-symmetric spaces and applied to the
$AdS_5\times S^5$ superstring in \cite{Hollowood:2014qma}.
}
It has been conjectured in \cite{Hollowood:2014rla,Hollowood:2014qma} that like the $\eta$-deformation, these also can be interpreted as a quantum-group deformation but in this case   the quantum group parameter is a root of unity $q=e^{i \pi/ k}$.

Although, at a first glance, the $\eta$- and $\lambda$-deformations may seem quite different since for
instance the corresponding $\s$-models have different isometry groups, they are, in fact, closely related.
At the level of currents, Rajeev observed some years ago \cite{Rajeev:1988hq}, that the canonical Poisson-structure of the PCM admits a one-parameter deformation which defines two commuting Kac--Moody algebras and preserves integrability. In the case of $SU(2)$, a brute force calculation in   \cite{Balog:1993es} led to a Lagrangian realization of Rajeev's canonical structure.   For arbitrary groups,  the $\eta$- and $\lambda$-deformations provide Lagrangian realisations for this Poisson-structure but for different ranges of Rajeev's deformation parameter. The connection between the $\eta$- and $\lambda$-deformations  is expected to be a bracket-preserving canonical transformation  followed by an appropriate analytic continuation of the deformation parameter and of the fields.    Specifically, the implementation of this transformation turns out to be a generalisation of T-duality known as Poisson--Lie (PL) T-duality  \cite{KS95a,KS95b} which can be understood as a canonical equivalence between a pair of $\s$-models \cite{Sfetsos:1996xj,Sfetsos:1997pi}.

PL T-duality incorporates the familiar Abelian T-duality and
non-Abelian T-duality as well as cases in which no isometries are present.  The crucial idea is that a $\sigma$-model
possess some currents for the action of a group $G$  that, although they are not conserved in the usual sense,
are covariantly conserved with respect to a dual group $\tilde{G}$.  The choice of groups $G$ and $\tilde{G}$ are
constrained such that the direct sum of the corresponding algebras defines a Drinfeld double $\frak{d}= \frak{g} \oplus \tilde{\frak{g}}$.

As shown for the case of principal chiral models in \cite{Klimcik:2002zj}, the YB $\sigma$-models take precisely the form of one-half of a PL T-dual related pair.   Recently this PL action has been considered in the case of symmetric spaces \cite{Vicedo:2015pna} where it was shown that it leads to an equivalence between the Hamiltonian of the YB $\sigma$-model on the {\em real} branch (when it can be defined as per footnote \ref{jsksls} and the $\lambda$-deformation.   This does not quite explain the link between the $\eta$-deformation and the $\lambda$-deformation since the former is on the {\em complex} branch.  Instead,  one should start with an $\eta$-deformation (i.e. a YB $\s$-model on the complex branch),   perform a  PL T-duality using the double $\frak{d}=\frak{g}^\mathbb{C}$ (the complexification of $\frak{g}$) and then analytically continue certain coordinates and the deformation parameter to arrive at the $\lambda$-deformation.  At the present moment the details of this process are not fully understood for an arbitrary group or coset.
Recently, a two-dimensional example based on $SU(2)/U(1)$ has been provided in \cite{Hoare:2015gda} and conjectured to hold in general.

\section{Summary and outlook}

The focus of this paper is to consider certain multi-parameter generalisations of both the $\eta$- and $\lambda$-deformations for which less in known.  In the case of $\eta$-deformations, there is an integrable class of two-parameter YB deformations introduced for an arbitrary semi-simple group in \cite{Klimcik:2008eq,Klimcik:2014}.  Such deformations are called bi-YB deformations.   For $\lambda$-deformations, the gauging procedure of  \cite{Sfetsos:2013wia} can be performed starting with an arbitrary coupling matrix in a PCM.  This gives rise to a wide family of deformations labeled by a matrix $\lambda_{ab}$. For an isotropic coupling, $\lambda_{ab}=\lambda \delta_{ab}$,   integrability was proven directly in  \cite{Sfetsos:2013wia} for an arbitrary semi-simple group, for symmetric cosets in \cite{Hollowood:2014rla} and for semi-symmetric spaces in \cite{Hollowood:2014qma}.  For the case of $SU(2)$, it has been shown in \cite{Sfetsos:2014lla} that providing $\lambda_{ab}$ is symmetric, the deformation is integrable. In this paper we will construct a two-parameter family of integrable $\lambda$-deformations in which $\lambda_{ab}$ acquires some
off-diagonal antisymmetric components.

There are quite a few novel results in this paper and so now is an opportune moment to summarise them:
\begin{itemize}
\item
We give concise expressions for the one-loop beta-functions for the deformation parameters of bi-YB deformations for a general group $G$.
These are parametrised by just two parameters out of the $(\dim G)^2$ possible ones. The fact that the flow preserves this two parameter truncation
renders the construction as non-trivial.

\item We construct a new class of two-parameter generalised $\lambda$-deformations that are obtained by performing a standard (one-parameter isotropic) $\lambda$-deformation on a YB deformed $\sigma$-model.  We construct a Lax pair representation for the equations of motion
    for groups as well as for symmetric cosets, hence demonstrating the integrability of the aforementioned deformations.
   \item For groups these multi-parameter deformations can be constructed in general but for
    general cosets  we find a stringent condition for them to be admissible. For compact groups an example of when this condition {\em is} satisfied  is given by   $\mathbb{CP}^2 = SU(3)/U(2)$ and example of when it {\em is not} satisfied is given by $S^5=  SO(6)/SO(5)$. 
   
\item  We study the connection between such generalised $\lambda$-deformations and the bi-YB $\eta$-deformation through examples based on the group $SU(2)$ and the coset $SU(2)/U(1)$.   We show that PL T-duality plus analytic continuation relate these deformations to a generalised $\lambda$-deformation of the type described above.

\item
For the $SU(2)/U(1)$ case we show that the $\lambda$-deformation, obtained after PL T-duality and
analytic continuation, is integrable by explicitly constructing the Lax pair. We also interpret the generalised $\lambda$-deformation as driven by para-fermion bilinears
of the exact $SU(2)/U(1)$ coset CFT.

\end{itemize}

The structure of the paper is as follows:  In section \ref{sec:biYB} we review the construction of the YB and
bi-YB $\eta$-deformations and their interpretation in terms of PL T-duality; in section \ref{sec:biYBrenorm}
we describe the one-loop renormalisability of the bi-YB deformation; in section \ref{sec:genlambda} we describe the
generalised $\lambda$-deformations and their integrability properties;  in \ref{sec:SU2groupexamples} we study
explicit examples based on $SU(2)$ and the $SU(2)/U(1)$ coset.

 This work suggests many exciting avenues for further related research. These include the embedding of our new $\l$-deformations into the full type-II string theory as well as applications in holography. The examples presented based on $SU(2)$ and $SU(2)/U(1)$  make the Poisson--Lie plus analytic continuation connection explicit between the two-parameter $\lambda$- and $\eta$-deformations; we expect this to hold in full generality.
 It will also be interesting to extend considerations of the generalised $\lambda$-deformations to semi-symmetric spaces. In this work we consider only classical integrability and understanding how this transfers to the quantum setting will be an important direction.

 \section{YB type models and Poisson--Lie T-duality}
\label{sec:biYB}

Before we begin let us set conventions that are used throughout.  For a compact semisimple Lie group $G$ 
corresponding to an algebra $\frak{g}$, we parametrise a group element $g\in G$ by local coordinates 
 $X^\m$, $\m=1,2,\dots , \dim(G)$. The right and left invariant Maurer--Cartan forms, as well as the orthogonal
matrix (or adjoint action) relating them, are defined as
\be
\begin{aligned}
  & L^{a}_{\pm} = L^a_\m \del_\pm X^\m  =  -i\, {\rm Tr}(T_a g^{-1}\del_\pm g )\ ,
 \quad R^a_\pm =  R^a_\m \del_\pm X^\m = -i\, {\rm Tr}(T_a  \del_\pm g g^{-1} ) \ ,
\\
 & R^a_\m = D_{ab}L^b_\m\ ,  \quad D_{ab}(g)={\rm Tr}(T_a g T_b g^{-1})\ .
 \label{jjd}
 \end{aligned}
\ee
The generators $T_a$ obey $[T_a,T_b]=i f_{ab}{}^{c} T_c$, are normalised as ${\rm Tr}(T_a T_b)=\d_{ab}$,  and with respect to the Killing metric, defined by $f_{ac}{}^{d}f_{bd}{}^{c}=-c_G\,\d_{ab}$,  the structure constants with lowered indices  $f_{abc}$ are totally antisymmetric. Group theoretic indices are frequently raised out by using $\delta_{ab}$.   World-sheet light cone coordinates are defined as $\s^{\pm} = \tau \pm \sigma$.

 \subsection{YB-type deformations of  Principal Chiral Models}

The bi-invariant (isotropic) PCM for the group $G$ is given by
\be
S_{ \textrm{PCM}}  = \frac{1}{2\pi t} \int_{\Sigma} \mathrm{d}^{2}\s  \,R_{+}^T R_{-} \ ,
\ee
in which $\Sigma$ is the world sheet and $t^{-1}$ is a dimensionless coupling, playing the r\^ole of  tension measured in units of $\alpha^\prime$,  that we shall need to keep track of  in what follows.   The PCM is classically integrable and its equations of motion can be readily recast in a Lax pair formulation.

Given a solution ${\cal R}$ of the modified YB equation \eqref{mYB},  the integrable YB deformation of this PCM is given by \cite{Klimcik:2002zj}
\be\label{eq:YBsigmamodel}
S_{ \textrm{YB} }  = \frac{1}{2\pi t} \int \mathrm{d}^{2}\s  \,R_{+}^T  (\mathbb{1} - \eta {\cal R} )^{-1} R_{-}  \ ,
\ee
{ where $\eta$ is a real non-negative constant.}
When ${\cal R}$ is restricted to be on the complex branch (i.e. $c^2<0$ in \eqref{mYB}) then we use the terminology $\eta$-deformation to refer to this model but for the time being we keep ${\cal R}$ general.
A two-parameter deformation, known as the bi-YB deformation, is given by \cite{Klimcik:2008eq,Klimcik:2014}
\be
\label{eq:biYBsigmamodel}
S_{\textrm{bi-YB} }  = \frac{1}{2\pi t} \int \mathrm{d}^{2}\s  \,R_{+}^T  (\mathbb{1} - \eta {\cal R} - \zeta  {\cal R}_g )^{-1} R_{-}  \ ,
\ee
where ${\cal R}_g = {\textrm{ad}}_g {\cal R}   {\textrm{ad}}_{g^{-1}} = D {\cal R} D^T$. Since both ${\cal R}$ and ${\cal R}_g$ are antisymmetric, this action is invariant under the parity transformation
\be
\s_+\leftrightarrow\s_-\,,\quad \eta\mapsto-\eta\,,\quad \zeta\mapsto-\zeta\,,
\ee
as well as the transformation
\be
g\mapsto g^{-1}\,,\quad \eta\leftrightarrow\zeta\ .
\ee

\no
It is convenient to consider a general action
 \begin{equation}
  \label{YBaction}
  S_{\eta,E} = \frac{1}{2\pi t} \int  \mathrm{d}^2 \sigma\,  R_+^T (E_g - \eta \R )^{-1} R_- \,,
\end{equation}
where $E$ is an arbitrary constant matrix and $E_g =   {\textrm{ad}}_g E {\textrm{ad}}_{g^{-1}} = D E D^T$.  This reduces to  \eqref{eq:YBsigmamodel} when $E= \mathbb{1}$
and to  \eqref{eq:biYBsigmamodel}  when $E= \mathbb{1} - \zeta {\cal R}$.   By interchanging right-invariant Maurer--Cartan forms with left ones using  eq.~\eqref{jjd}, this action can be rewritten as
  \begin{equation}
  \begin{aligned}\label{eq:PLform}
  S_{\eta,E}
   &= \frac{1}{2\pi t} \int \mathrm{d}^2 \sigma L_+^T ( E-  \eta   \R_{g^{-1}})^{-1} L_- \\
   & = \frac{1}{2\pi t\,\eta}  \int  \mathrm{d}^2 \sigma L_+^T (M - \Pi )^{-1} L_-\,,
  \end{aligned}
\end{equation}
  where
  \begin{equation}
\label{eq:pidef}
M = \frac{1}{\eta} E - \R \ , \quad \Pi = \Pi(g) = \R_{g^{-1}}-\R\ .
\end{equation}
This rewriting of the action exposes an important property; it has a left acting PL symmetry
 \cite{KS95a,KS95b}.
Although \eqref{eq:PLform} is not invariant under the left action of $G$, the currents ${\cal J}_a$ corresponding to this left action
obey the modified conservation law\footnote{
The world-sheet coordinates $(\s^+,\s^-)$ and $(\tau,\sigma)$ are related by
\begin{equation*}
\s^\pm: = \tau \pm \s \ ,\qq
\del_0 :=\del_\tau= \del_+ +\del_- \ ,\quad \del_1:=\del_\s = \del_+ - \del_- \ ,
\end{equation*}
so that $\star\,\mathrm{d}\s^\pm =\pm \mathrm{d}\s^\pm\, \&\, \star \mathrm{d}\tau = \mathrm{d}\s\ , \star\,\mathrm{d}\sigma = \mathrm{d}\tau$ in Lorentzian signature.
}
\begin{equation}
\begin{split}
\mathrm{d} \star {\cal J}_a = \frac{\eta}{2} \tilde{f}^{bc}{}_a \star {\cal J}_b \wedge \star {\cal J}_c \ ,
\end{split}
\end{equation}
where
\be\label{eq:ftilde}
\begin{split}
&{ {\cal J}_+=(E_g^T+\eta\cR)^{-1}\,R_+\,,\quad  {\cal J}_-=(E_g-\eta\cR)^{-1}\,R_-\,,}\\
& \tilde f^{ab}{}_c= - \R_{d}{}^a f_{dc}{}^b+ \R_{d}{}^b f_{dc}{}^a =-\tilde f^{ba}{}_{c}\ ,
\end{split}
\ee
with no symmetry at the third index $c$.
Algebraically, the $\tilde{f}$ are the structure constants that arise from a second Lie-algebra $\frak{g}_{\cal R}$  defined by the bracket
\be
[A,B]_{\R}=[\R A,B]+[A,\R B] \ ,  \quad \forall  A,B \in \frak{g} \, .
\ee
Thus over the vector space of  $\frak{g}$ we have two algebras, $\frak{g}$ and $\frak{g}_{\cal R}$ whose direct sum defines a Drinfeld double
$\frak{d} = \frak{g}\oplus  \frak{g}_{\cal R}$  (see  appendix \ref{Rappendix} for details). One needs here to distinguish a little between the complex branch ($c^2<0$ in the mYB \eqref{mYB})
for which the Drinfeld double is the complexification $\frak{d}= \frak{g}^\mathbb{C}= \frak{g}\otimes \mathbb{C}$ of a real Lie-algebra ${\frak g}$,  and the real branch ($c^2>0$) in which case the double is given by  $\frak{d}= \frak{g}^{diag} \oplus \frak{p}$  (further discussion of the construction of the Drinfeld double from the $\R$-matrix can be found in \cite{Vicedo:2015pna} and also chapter 4 of \cite{bookBBT}).\footnote{Although it will not be discussed here, the utility of the classical branch ($c^2=0$) in describing integrable deformations was shown in \cite{Kawaguchi:2014qwa}  and the link to a wide class of known deformations including the gravitational duals of non-commutative Yang--Mills and Schr\"odinger deformations was elucidated in \cite{Matsumoto:2014gwa,Matsumoto:2015uja} and \cite{vanTongeren:2015soa,vanTongeren:2015uha}.}

When a $\sigma$-model is invariant under some action of a group $G$ then one can dualise the theory; when the group is Abelian, this is just T-duality and when the group is non-Abelian this leads to so-called non-Abelian T-duality.   Although   not invariant under left action of $G$, the PL symmetry of  eq.~\eqref{eq:PLform} is such that there is still a generalised notion of T-duality that is applicable.  This goes by the name of PL T-duality  which is an equivalence between   two $\s$-models\footnote{To match the conventions of \cite{Valent:2009nv}:
$g\mapsto g^{-1}\,, \Pi(g)\mapsto -\Pi_{KL}(g^{-1})$
and of \cite{Sfetsos:1999zm}:
$M\mapsto-\mathrm{E}^{-1}_0\,, t\mapsto-t$\,.}
\be\label{eq:PLTdualpairs}
\begin{aligned}
&S[g] =\frac{1}{2\pi t\,\eta} \int  \mathrm{d}^2 \sigma L_+^T (M - \Pi )^{-1} L_-\,, \quad g \in \frak{g} \, ,   \\
&\tilde{S}[\tilde{g}] =\frac{1}{2\pi t\,\eta}  \int \mathrm{d}^2 \sigma \tilde{L}_+^T (M^{-1} - \tilde\Pi )^{-1} \tilde{L}_-\, , \quad  \tilde g \in \frak{\tilde g} \, .
  \end{aligned}
\ee
The matrix $M$ was defined in \eqref{eq:pidef} but can, for the purposes of dualisation, be an arbitrary constant
matrix.\footnote{Though, of course, for an arbitrary choice of M  the theory is not expected to be integrable.}  Here the
algebras $\frak{g}$ and $ \frak{\tilde g}$, with generators $T_a$ and $\tilde{T}^a$, form a Drinfeld
double $\frak{d}= \frak{g}\oplus  \frak{\tilde g}$ which is equipped with an inner product such
that  $\langle T_a , T_b \rangle = \langle \tilde T^a , \tilde T^b \rangle= 0 $ and    $\langle T_a , \tilde T^b \rangle = \delta_a^b$.
 The group theoretic matrix $\Pi$ is defined by
\be
a_a{}^b = \langle g^{-1} T_a g , \tilde{T}^b \rangle \ , \quad  b^{ab} = \langle g^{-1} \tilde T^a g , \tilde{T}^b \rangle \ , \quad \Pi = b^T a\,,
\ee
with similar for the tilded quantities.
In a following section, we study in detail the case of $\frak{g} = \frak{su}(2)$, for a YB deformation on the complex branch where the
relevant Drinfeld double is $D=\frak{g}^\mathbb{C}= \frak{su}(2)\oplus \frak{e}_3$.

As a final remark in this section we note that the bi-YB deformation is neither left nor right invariant under the action of $G$ but
instead is {\em both} left and right PL symmetric. We will however, in this work, only consider PL T-duality applied to the left PL symmetry.

 \subsection{YB-type deformations of symmetric coset spaces}
 \label{lapp}

To introduce YB-type deformations on symmetric cosets, let us first take a small digression and consider the construction of
general PL-type theories on cosets originally considered in the literature in \cite{Klimcik:1996np,Sfetsos:1999zm}
and revisited in \cite{Squellari:2011dg}. Here we follow
the construction of  \cite{Sfetsos:1999zm} which turns out to be relevant for our purposes.
Consider the general form of PL T-dual
pairs  given in  eq.~\eqref{eq:PLTdualpairs} which {\em a priori} describe the dynamics of $\dim G$ degrees of freedom.
At certain special points in the moduli space of such theories, i.e. for particular choices of $M$, the theory may develop
a gauge invariance under some subgroup $H \subset G$ leaving a dynamical theory for just $\dim G/H$ coset degrees of freedom.
 Let us use the notation in which $T_i$ are generators of the sub-algebra $\frak{h}$  corresponding to the subgroup $H$ and $T_\a$
are the remaining generators of $\frak{k}=\frak{g}-\frak{h}$.  Points  where one might expect to find an enhanced $H$ gauge redundancy can be
reached by taking a limit, understood be acting uniformly on all matrix elements,
\be
M \big{|}_{\frak{h}} := M_{ij}  \rightarrow 0  \ .
\ee
In this limit one can see that the first of \eqref{eq:PLTdualpairs} becomes \cite{Sfetsos:1999zm}
\be \label{eq:PLcoset}
S[g] \mapsto \frac{1}{2\pi t\,\eta} \int  \mathrm{d}^2 \sigma L_+^\a \Sigma_{\a\b}  L_-^\b \,, \quad   \Sigma_{\a\b}  =(M_{\a\b} - \Pi_{\a \b})^{-1}  \ .
\ee
Although the action only involves the left-invariant one-forms in the coset, it is still not a coset theory.
This will be the case if it develops gauge invariance under the action of the subgroup.  The condition on invariance is given in \cite{Sfetsos:1999zm} as
\be \label{eq:gaugeinv}
\tilde{f}^{\a\b}{}_i  = f_{i \g}{}^{\a} M^{\g\b} + f_{i\g}{}^{\b} M^{\a \g}  \ .
\ee
{Having understood this general construction one can now forget this limit procedure and simply look for $\dim G/H$ matrices that obey  eq.~\eqref{eq:gaugeinv} with which to build a coset theory.  To do so we use as a starting point the matrix $M$ corresponding to the bi-YB $\s$-model given in eq.~\eqref{eq:pidef}
and define the $\dim G/H$ matrix $M_{\a\b}$ entering into  eq.~\eqref{eq:PLcoset} as the projection into the coset
\be
M_{\a\b}  =  \left(\frac{1}{\eta } \mathbb{1} - \left( \frac{\zeta}{\eta } +1  \right) {\cal R} \right)_{\a \b}  \ .
\ee
Making use of the expressions for the dual structure constants in terms of the ${\cal R}$-matrix given in eq. \eqref{eq:ftilde}, the gauge invariance condition becomes
\be
\label{jdsnsd}
0 = \frac{\zeta}{\eta} \left(- \R_{\g}{}^{\a} f_{\g i}{}^{\b}+ \R_{\g}{}^{\b} f_{\g i }{}^{\a}    \right) =  \frac{\zeta}{\eta}  \tilde{f}^{\a\b}{}_i  \ .
\ee

For $\zeta \neq 0$ this implies that the projection of the ${\cal R}$-bracket into the sub-algebra $\frak{h}$ must vanish
\begin{equation}
\label{coset.constr.bracket1}
[ X , Y]_{\cal R} |_{\frak{h}} = ( [{\cal R}X , Y]+   [X , {\cal R} Y]|)  |_{\frak{h}}   = 0 \ ,\quad \forall\, X,Y\, \in \frak{k} \ .
\end{equation}

For the case of $\zeta=0$,  eq.~\eqref{eq:PLcoset} with this choice of $M_{\a\b}$   defines a theory that for a symmetric space (i.e.  $[\frak{h},\frak{h}] \subseteq \frak{h}$,
$[\frak{h}, \frak{k}]\subseteq \frak{k}$ and $  [\frak{k}, \frak{k}]\subseteq \frak{h}$) has  already been considered in the literature -  an integrable YB $\sigma$-model is given by  \cite{Delduc:2013fga}
\be\label{eq:YBsigmamodelCoset}
\begin{split}
 S_{\textrm{YB}, G/H } =   \frac{1}{2\pi t} \int \mathrm{d}^{2}\s  \,( P_1 L_{+})^T  (\mathbb{1}  - \eta {\cal R}_{g^{-1}}P_1 )^{-1}  P_1 L_{-}  \ ,\\
\end{split}
\ee
in which, to make contact with formulations given elsewhere, we introduce the projector $P_1$   into the coset generators $\frak{k}$ of the algebra  $\frak{g} = \frak{h} + \frak{k}$ so that $M_{\a\b} = (P_1 M P_1^T)_{\a\b}$.   The corresponding expression for the bi-YB case is given by
\be\label{eq:biYBsigmamodelCoset}
\begin{split}
S_{\textrm{bi-YB},G/H }    = \frac{1}{2\pi t} \int \mathrm{d}^{2}\s  \,( P_1 L_{+})^T  (\mathbb{1}- \zeta {\cal R}\,P_1  -
\eta {\cal R}_{g^{-1}}P_1 )^{-1} P_1 L_{-}  \ ,
\end{split}
\ee
in which one needs to also impose eq.~\eqref{coset.constr.bracket1}.
 Being a natural extension to cosets of the bi-YB deformations it will be rather interesting to study the integrability properties of this theory.\footnote{We that Beno\^it Vicedo for communications on this point.}  This theory admits a PL T-dual and it seems likely to us that it can be related, in general, to certain 2-parameter integrable $\lambda$-deformations that will be constructed in section \ref{sec:symcoset} via a such a duality plus analytic continuations -- this will be illustrated with an example in the rank 1 case.  We leave a more direct study of the theory defined by eq.~\eqref{eq:biYBsigmamodelCoset} for the future. However at this stage it is worth studying the constraint  of eq.~\eqref{coset.constr.bracket1} in more detail.   It is certainly a stringent condition on the admissibility of the subgroup $H\subset G$ given an ${\cal R}$-matrix.  It is rather trivial to see that this constraint can be solved in the simple rank 1 case but the existence of solutions for general $G/H$ is less clear. 
\\
The symmetric coset  $\mathbb{CP}^2 = \frac{SU(3)}{U(2)}$  provides  an example in which eq.~\eqref{coset.constr.bracket1} {\em is} satisfied.   In the usual $SU(3)$ Gell-Mann basis of generators $\l_a$, $a=1, \dots, 8$ the anti-symmetric ${\cal R}$-matrix solving the mYB for $c^2 < 0$ acts as that
\begin{equation*}
{\cal R} \circ \{ \l_2, \l_5 , \l_7 \} \mapsto \{ \l_1,  \l_4 , \l_6 \}\,,
\end{equation*}
with vanishing action on the rest of the generators.
Choosing the $U(2)$ subgroup that generated by $\{\l_1, \l_2, \l_3 ; \l_8\}$ one finds that  eq.~\eqref{coset.constr.bracket1} holds. On the other hand, for the $U(2)$ subgroup generated by $\{\l_4, \l_5, \frac{1}{2}\l_3+\frac{\sqrt{3}}{2}\l_8;     \frac{\sqrt{3}}{2}\l_3 -\frac{1}{2}\l_8 \}$ -- which also defines a symmetric space -- the constraint  eq.~\eqref{coset.constr.bracket1} {\em does not} hold.

For $S^5 = SO(6)/SO(5)$ the anti-symmetric ${\cal R}$-matrix  solving the mYB for $c^2 < 0$ acts as
\begin{equation*}
{\cal R} \circ   \{T_{23}, T_{24}, T_{25}, T_{26}, T_{45},T_{46} \} \mapsto \{T_{13},  T_{14} ,T_{15}  , T_{16}, T_{35}, T_{36}  \} 
\end{equation*}
with vanishing action on the rest of the generators. Picking an explicit basis for the subgroup $T_{ij}$ for $i<j  = 1,\dots, 5$ one finds that eq.~\eqref{coset.constr.bracket1} {\em does not} hold.

\subsubsection*{General YB-type cosets}

Following the procedure described above we can project the general "YB-type" of models, \eqref{YBaction} or equivalently \eqref{eq:PLform}, on a general coset $G/H$.
The end result of this procedure reads
\be
S_{\eta,E,G/H} = \frac{1}{2\pi t} \int  \mathrm{d}^2 \sigma\,  ( P_1 L_{+})^T (E P_1 -\eta {\cal R}_{g^{-1}}P_1 )^{-1} P_1 L_{-}  \,,
\ee
whereas the gauge invariance condition reads
\be
E_{\a\g}\,f_{\b\g i}+E_{\g\b}\,f_{\a\g i}=0\,.
\ee
This condition is trivially satisfied in the YB case whereas for the bi-YB case, $E=\mathbb{1}-\zeta\cR$, and \eqref{jdsnsd} and \eqref{eq:biYBsigmamodelCoset} trivially follow.
We emphasise that this coset construction for generic $G/H$ and $E$ is not integrable.

\section{Quantum aspects of the bi-YB model}
\label{sec:biYBrenorm}

The renormalisability of the general PL T-dual $\sigma$-models in \eqref{eq:PLTdualpairs} at one-loop was proved in \cite{Valent:2009nv}.
In \cite{Sfetsos:2009dj} it was demonstrated that the one-loop RG flow matches one obtains for the coupling matrices $M_{ab}$ from
both of the dual theories are in fact equivalent which is physically sensible given the canonical equivalence of PL related
$\s$-models \cite{Sfetsos:1996xj,Sfetsos:1997pi}.
This can also be understood in terms of a first-order duality invariant type formalism \cite{KS95a} (c.f. the doubled formalism of abelian T-duality) from which the one-loop beta functions for the couplings contained in the matrix $M_{ab}$ of eq.~\eqref{eq:PLTdualpairs}  can be obtained \cite{Sfetsos:2009vt}.   Although the full expressions for the remormalisation of $M$ are rather involved, here we are able to specialise to the case of the bi-YB equation and obtain a very simple set of RG equations governing the flow of the deformation parameters $\eta$ and $\zeta$.

One should emphasise that although the most general $\sigma$-model with $M_{ab}$ encoding $(\dim G)^2$ coupling constants is renormalisable, this does not imply that the renormalisability of the bi-YB $\sigma$-model.  The later is a truncation parametrised by just two parameters out of the $(\dim G)^2$ possible ones. The RG flow equations for $M_{ab}$ could, in principal, not preserve this truncation.    The fact that the flow preserves this two parameter truncation renders the construction as non-trivial.

 Before specialising to the bi-YB case, we first present the general RG equations for the models of \eqref{eq:PLTdualpairs}  which we shall do using   the notation introduced in \cite{Sfetsos:2009dj}.  We define
\begin{equation}
A^{ab}{}_{c} = \tilde f^{ab}{}_c - f_{cd}{}^a M^{db}\ ,\qq
B^{ab}{}_{c} = \tilde f^{ab}{}_c + M^{ad}f_{dc}{}^b \ ,
\end{equation}
as well as their duals
\begin{equation}
\tilde A_{ab}{}^{c} =  f_{ab}{}^c - \tilde f^{cd}{}_a M^{-1}_{db}\ ,\qq
\tilde B_{ab}{}^{c} = f_{ab}{}^c + M^{-1}_{ad}\tilde f^{dc}{}_b \ .
\end{equation}
Using these we construct
\begin{equation}
\label{RL.PLT}
\begin{split}
&L^{ab}{}_c  =   \ha [M_s^{-1}]_{cd}\left( B^{ab}{}_e M^{ed} + A^{db}{}_e M^{ae}- A^{ad}{}_eM^{eb}
  \right) \ ,\\
&R^{ab}{}_c  =  \ha [M_s^{-1}]_{cd}\left( A^{ab}{}_e M^{de} + B^{ad}{}_eM^{eb} - B^{db}{}_e M^{ae}
\right) \,, \\
&\tilde L_{ab}{}^c  =   \ha [\tilde M_s^{-1}]^{cd}\left(
\tilde B_{ab}{}^e M^{-1}_{ed} + \tilde A_{db}{}^e M^{-1}_{ae}- \tilde A_{ad}{}^e M^{-1}_{eb}
  \right) \ ,\\
&\tilde R_{ab}{}^c  =  \ha [\tilde M_s^{-1}]^{cd}\left( \tilde A_{ab}{}^e M^{-1}_{de}
+ \tilde B_{ad}{}^e M^{-1}_{eb} - \tilde B_{db}{}^e M^{-1}_{ae}
\right) \ ,
\end{split}
\end{equation}
where
\begin{equation}
M_s = \ha (M+M^T) \ ,\quad\tilde M_s = \ha \left[M^{-1} + M^{-T}\right]\ .
\end{equation}
The one-loop RG flows are
\be
\label{RG1}
\frac{1}{t\, \eta}\,  (M^{ab})^\cdot= R^{ac}{}_d L^{db}{}_c+R^{ab}{}_c\xi^c\,,
\ee
and
\be
\label{RG2}
\frac{1}{t\, \eta}\, (M^{-1}_{ab})^\cdot= \tilde R_{ac}{}^d \tilde L_{db}{}^c+\tilde R_{ab}{}^c\tilde\xi_c\ ,
\ee
where $\xi^c,\tilde\xi_c$ are constants corresponding to field redefinitions (diffeomorphisms)
and dot corresponds to derivatives with respect to the logarithmic energy scale. It was shown in \cite{Sfetsos:2009dj} that
the two systems \eqref{RG1} and \eqref{RG2} turn out to be equivalent.\footnote{The diffeomorphism terms were not incorporated in the analysis of  \cite{Sfetsos:2009dj} but are  easily included by relating them via $\tilde \xi_a=-M^{-1}_{ab}\,\xi^b$ and using the identity, proved in   \cite{Sfetsos:2009dj},  $R^{ab}{}_{c} = M^{a e} M^{f b} M_{g c}^{-1} \tilde{R}_{ef}{}^g$ .}

We can now specialise these relations to the bi-YB deformation for which we have the corresponding $M$ matrix
\be\label{eq:MbiYB}
M=\frac1\eta\left(\mathbb{1}-(\eta+\zeta)\cR\right)\,.
\ee
Making heavy use of the identities obeyed by ${\cal R}$ detailed in the appendix, see eqs.~\eqref{mYB1} and \eqref{mYB2}, one finds that
\be
\label{RL.biYB}
\begin{split}
&L^{ab}{}_c=-\frac\zeta\eta\,\cR_{am}f_{bmc}+\frac{c^2(\eta^2-\zeta^2)-1}{2\eta}\,f_{abc}\,,\\
&R^{ab}{}_c=\frac\zeta\eta\,\cR_{bm}f_{amc}-\frac{c^2(\eta^2-\zeta^2)-1}{2\eta}\,f_{abc}\, ,
\end{split}
\ee
in which we recall that $c^2$ is the parameter entering into the mYB equation.   To ensure that the renormalisation of $M$ stays within the truncation specified by eq.~\eqref{eq:MbiYB} one requires   a redefinition generated by   $\xi^a=-\zeta/\eta\,f^{abc}\,\cR_{bc}$ in \eqref{RG1}.  Upon plugging \eqref{RL.biYB} into \eqref{RG1} and making further use of the identities \eqref{mYB1} and \eqref{mYB2}
and the Jacobi identity, one then finds the system of one-loop RG equations for $\zeta, \eta$ and $t$ given by
\be
\label{RG.biYB}
\boxed{\begin{split}
&\dot \eta=\frac{c_G\,t\,\eta}{4}\,\left(1-c^2(\eta-\zeta)^2\right)\,\left(1-c^2(\eta+\zeta)^2\right)\,,\\
&t\,\eta\, {\rm and}\,\,\, \zeta/\eta=\rm{constants}\,.
\end{split}}
\ee
For the corresponding $\eta$-deformation we set $\zeta=0$ and again $t\,\eta={\rm constant}.$ In this particular case and for
$c^2=-1$ the $\b$-functions were derived in \cite{Squellari:2014jfa}. However, in that work the ratio $t/\eta$ was found to be constant, a statement
with which we disagree.  In fact, it turns out to be rather important that it is the combination $t \eta$ that is a RG invariant; under the Poisson--Lie plus analytic continuation that relates $\eta$-type deformations to $\lambda$-type we require that $ 4 t\, \eta = i k^{-1}$ where $k$ is quantised WZW level that should not run.

\no
We elaborate briefly on the form of the solution of the $\beta$-function for the $\eta$-deformed theory (setting for the moment $\zeta=0$). When $c^2<0$
then it is evident that the energy scale is a bounded function of the coupling $\eta$, which implies that in this model the UV and the IR
energy regimes cannot be reached. {Consider for example $c=i$, then the RG flows \eqref{RG.biYB}, setting $\zeta=0$, can be easily integrated
\be
\label{RG.YB.Integrated}
\frac{c_G\,t\,\eta}{2}\ln\frac{\mu}{\mu_0}=\frac{\eta}{1+\eta^2}+\tan^{-1}\eta\,,\quad t\,\eta={\rm constant}\,,\quad
\eta\in[0,\infty)\,,
\ee
where $\mu_0$ is an integration constant. We   note that the right hand side of \eqref{RG.YB.Integrated} 
is bounded on the domain $[0,\pi/2)$.  $\eta$ will thus diverge at some UV scale $\frac{\mu_{UV}}{\mu_0} = e^{\frac{\pi}{c_G t\,\eta}}$ and achieves its minimum value $\eta \rightarrow 0$ at an IR scale $\mu_0$. One should be careful; thinking of the theory as a non-linear $\sigma$-model means we should only trust perturbative results (including the above $\beta$-functions) when the curvature radius of the target space is small compared to $t\, \eta $.  However in the limits of $\eta \rightarrow 0 $ and $\eta \rightarrow \infty$ this is no longer the case and perturbation theory breaks down.\footnote{{ For example consider the scalar curvature of the YB $\sigma$-model for $su(2)$ (whose target space is just a squashed sphere): 
\begin{equation*}
R=\frac{t\,\eta}{2}\,\frac{(3-\eta^2)(1+\eta^2)}{\eta}\,,
\end{equation*}
and so the perturbative description in powers of $t\,\eta$ breaks down as $\eta\to0$ and as $\eta\to\infty$.}
}}  In contrast, when $c^2>0$ {($c^2=0$)} it is easy to see, {via an analogue to the above integration, that in the UV $\eta \rightarrow 1^-$ ($\eta\rightarrow\infty$) and  $\eta \rightarrow 0^+$ at an IR scale $\mu_0$.}

\section{Generalised integrable $\lambda$-deformations}
\label{sec:genlambda}

The purview of this section is to introduce a generalised notion of $\lambda$-deformations and to show for a particular case, which can be thought of as the $\lambda$-deformed YB $\sigma$-model, classical integrability is ensured through the existence of spectral dependent classical Lax pair.

 \subsection{Constructing the deformation}
\no
This subsection reviews the construction of $\l$-deformations by following the original literature \cite{Sfetsos:2013wia} and also \cite{Sfetsos:2014lla}.
Compared to that work we have  formulated the PCM and the WZW model  in terms of the right invariant Maurer--Cartan forms so as to match the YB $\s$-models of the previous section.

We begin with PCM on the group manifold for an element $\hat{g}\in G$ but generalised to incorporate an arbitrary, not-necessarily bi-invariant, constant matrix $\hat{E}_{ab}$,
\be
\label{eq:PCM2}
S_{{\rm PCM}}(\hat{g}) = \frac{1}{2\pi} \int \mathrm{d}^2\sigma\hat{E}_{ab} R_{+}^{a}(\hat{g})  R_{-}^{b}(\hat{g})\,.
\ee
We also consider a WZW action for a group element $g  \in G$    defined by
\be
\label{eq:WZW}
S_{{\rm WZW},k}(g) =
\frac{k}{4\pi} \int_{\Sigma}  \mathrm{d}^2\sigma R^a_+ R^a_- \,
  -\frac{k}{24\pi}\int_{{\cal B}} f_{abc}
  R^a\wedge R^b\wedge R^c\ ,
\ee
where ${\cal B}$ is an extension such that $\partial {\cal B} = \Sigma$ and the normalisation is such that, with our conventions for the generators, $k$ is an integer for $SU(N)$.
The approach of \cite{Sfetsos:2013wia} was to consider the sum of the actions in \eqref{eq:PCM2} and \eqref{eq:WZW}
and to gauge a subgroup of the global symmetries that acts as
\be
\hat{g} \mapsto \hat{g}\, h \ , \quad g\mapsto h^{-1} g h \ , \quad h\in G \ .
\ee
This is achieved by introducing a connection $A= A^aT_a$ valued in the algebra of $G$ that transforms as
\be
A \mapsto   h^{-1} A h + h^{-1} \mathrm{d}h \ .
\ee
We  replace derivatives in the PCM with covariant derivatives defined as:\\ $D \hat{g}  =  \mathrm{d} \hat{g} -\hat{g}A$
and replace the $WZW$ with the $G/G$ gauged $WZW$ given by
\begin{equation*}
S_{{\rm gWZW},k}= S_{{\rm WZW},k} +   {k\ov 2\pi} \int  {\rm Tr}( A_+ \del_-  g  g^{-1}- g^{-1}\del_+ g A_- + A_+ g A_- g^{-1}- A_+A_-) \ .
\end{equation*}
The gauge symmetry can now be gauged fixed by setting $\hat{g}= \mathbb{1} $
such that all that remains of the gauged PCM is a quadratic term in the gauge fields.
The gauge fields, which are non-propagating, obey constraint type equations,
\begin{equation}\label{eq:gaugeconstraints}
( \lambda^{-1} - D ) A_- = iR_- \ , \quad ( \lambda^{-T} - D^T  )   A_+ = -i L_+  \   ,
\end{equation}
where we have introduce a generalised $\lambda$-deformation matrix,
\be\label{eq:lambdadef}
\quad  \l^{-1} = k^{-1}(\hat{E}+k\,\mathbb{1} ) \ .
\ee
Upon integrating out these gauge fields one finds the $\s$-model action \cite{Sfetsos:2013wia}
\be
S_{k,\l}(g) =
S_{{\rm WZW},k}(g) + {{k}\ov 2\pi}  \int  \mathrm{d}^{2}\s\,  L_+^a
(\l^{-1}-D)^{-1}_{ab}R_-^b  \ .
\label{tdulalmorev2}
\ee

Although the equations of motion of $g$ that arise from this action will be rather intricate it was shown in  \cite{Sfetsos:2014lla}
that when written in terms of the gauge fields obeying \eqref{eq:gaugeconstraints} they take a simpler form
\be
\label{eom.A}
\begin{split}
&\del_+A_--\del_-\left(\lambda^{-T}A_+\right)=[\lambda^{-T}A_+,A_-]\,,\\
&\del_+\left(\lambda^{-1}A_-\right)-\del_-A_+=[A_+,\lambda^{-1}A_-]\ .
\end{split}
\ee
Note that unless $\lambda = \mathbb{1}$ these are not conditions for a flat connection.

To prove integrability we would like to rewrite these equations of motion as a Lax equation
\be
\label{Lax.equation}
\mathrm{d}{\cal L}={\cal L}\wedge {\cal L}\quad {\rm or}\quad
\del_+{\cal L}_--\del_-{\cal L}_+=[{\cal L}_+,{\cal L}_-]\,,
\ee
where ${\cal L}_\pm={\cal L}_\pm(\tau,\sigma;\mu)$ depends on a spectral parameter $\mu\in\mathbb{C}$. For a general choice of $\hat E_{ab}$ one certainly would not expect this to be possible, thus posing an interesting question; for what choices of $\hat  E_{ab}$ is this an integrable system?

As a warm up let us revisit the isotropic case $\displaystyle \hat  E={1\ov t}\,\mathbb{1}$ which is known to be integrable
\cite{Balog:1993es,Sfetsos:2013wia} (see also \cite{Hollowood:2014rla,Itsios:2014vfa}).
In this case $\lambda = \lambda_0 \mathbb{1}$ and the equations of motion \eqref{eom.A} reduce to
\be
\del_\pm A_\mp=\pm\frac{1}{1+\l_0}\,[A_+,A_-]\,,\quad \lambda_0=\frac{k\,t}{1+k\,t}\, .
\ee
A Lax connection encoding these equations is given by
\be
{\cal L}_\pm=\frac{2}{1+\l_0}\,\frac{\mu}{\mu\mp1}\,A_\pm\,,\quad \mu\in\mathbb{C}\, .
\ee

\subsection{Generalisation to YB $\s$-models}

  The key idea in constructing integrable $\lambda$-deformations is to take two integrable theories (e.g. the bi-invariant (isotropic) PCM   together with the WZW) and reduce half of the degrees of freedom in such a way that what is left remains integrable. In order to find other examples where the generalised $\lambda$-deformation is integrable, it is natural to consider as a starting point  PCM's  \eqref{eq:PCM2} that are known to be integrable and then apply the $\lambda$-deformation.

\subsubsection{The group case}
 \label{sec:lambdaonYB}

We recall that the integrable YB $\sigma$-model defined in \eqref{eq:YBsigmamodel} can be written as a PCM of the form in \eqref{eq:PCM2} for the choice
\begin{equation}
\hat{E} = \frac{1}{\tilde t} (\mathbb{1} - \tilde\eta {\cal R})^{-1} \ ,
\end{equation}
so let us consider this as a starting point for a generalised $\lambda$-deformation.

With this choice of $\hat{E}$ one finds, making use of the mYB equation \eqref{mYB},  that the equation of motion \eqref{eom.A} admits the nice rewriting
\be
\pm\del_\pm \widetilde A_\mp=\tilde\eta[\cR \widetilde A_\pm,\widetilde A_\mp]+a\,[\widetilde A_+,\widetilde A_-]\, ,
\ee
where we have defined $\widetilde A_\pm=(\mathbb{1}  \pm\tilde\eta\,\cR)^{-1}\,A_\pm$ and
\be
a=\frac{1+c^2\tilde\eta^2\lambda_0}{1+\lambda_0}\ , \qq  \lambda_0=\frac{k\,\tilde t}{1+k\,\tilde t}\,.
\ee
From this rewriting one can then see that the equations of motion can be written in terms of a Lax connection as
\be
\boxed{
\label{LaxYB}
\begin{split}
&{\cal L}_\pm=(\a_\pm\mathbb{1}\pm\tilde \eta\,{\cal R})(\mathbb{1}\pm\tilde\eta\,{\cal R})^{-1}\,A_\pm\,,\\
&\a_\pm=\a_1+\a_2\,\frac{\mu}{\mu\mp1}\,,\quad \mu\in\mathbb{C}\,,\\
&\a_1=a-\sqrt{a^2-c^2\tilde\eta^2}\,,\quad \a_2=2\sqrt{a^2-c^2\tilde\eta^2}\, .
\end{split}
}
\ee
This result proves that for an arbitrary choice of group, in addition to the $\lambda$-deformation of the isotropic PCM, the $\lambda$-deformation of the YB $\sigma$-model is integrable.   This provides a two-parameter family of deformations  labelled by $\tilde\eta$ and $\lambda_0$.   We will see later for the specific case of $G=SU(2)$ that this two-parameter family can also be obtained as the PL T-dual combined with analytic continuation of the bi-YB deformation (on the complex branch).  We conjecture that such a relation holds true in general.

\subsubsection{The symmetric coset case}
\label{sec:symcoset}
Let us now consider applying these ideas to symmetric cosets.   Motivated by the integrability of the YB $\sigma$-model \eqref{eq:YBsigmamodelCoset} on a symmetric space $G/H$, corresponding to a Lie-algebra $\frak{g}= \frak{h}+\frak{k}$,    let us consider starting with the following
\be\label{eq:symcosetdata}
\hat{E}=\hat{E}_H \oplus\hat{E}_{G/H}\,,\quad\hat{E}_H=0\,,\quad \hat{E}_{G/H}=\frac{1}{\tilde t} (\mathbb{1} - \tilde\eta {\cal R})^{-1}.
\ee
Here ${\cal R}$ is an anti-symmetric matrix of dimension $\dim G - \dim H$ which one could--but need not--think of as the ${\cal R}$-matrix satisfying the mYB equation projected into the coset.  With this choice of $\hat{E}$ and assuming that the coset is a symmetric space,  the equations of motion \eqref{eom.A} simplify to
\ba
&& \del_\pm\widetilde B_\mp=-[\widetilde B_\mp,A_\pm]\,,\quad \del_\pm\left(\cR\widetilde B_\mp\right)=-[\cR\widetilde B_\mp,A_\pm]\, ,
\\
&& \del_+A_--\del_-A_+=[A_+,A_-]+\frac{1}{\l_0}[\widetilde B_+,\widetilde B_-]+\tilde\eta[\R\widetilde B_+,\widetilde B_-]-
\frac{\tilde\eta}{\l_0}[\widetilde B_+,\R\widetilde B_-]-
\tilde\eta^2[\cR\widetilde B_+,\R\widetilde B_-]\ ,
\nonumber\\
&& \del_+A_--\del_-A_+=[A_+,A_-]+\frac{1}{\l_0}[\widetilde B_+,\widetilde B_-]+\frac{\tilde\eta}{\l_0}[\R\widetilde B_+,\widetilde B_-]-
\tilde\eta[\widetilde B_+,\R\widetilde B_-]-
\tilde\eta^2[\cR\widetilde B_+,\R\widetilde B_-]\ ,
\nonumber
\ea
where we defined
\be
\widetilde B_\pm=(\mathbb{1}\pm\tilde\eta\R)^{-1}B_\pm\,,\quad
\lambda_0=\frac{k\,\tilde t}{1+k\,\tilde t}\,,\quad A_\pm\in \frak{h} \,,\quad B_\pm\in\frak{k}  \,.
\ee
For consistency of the two forms of equations of motion for $A_\pm$, one finds that the projection of the ${\cal R}$-bracket into the sub-algebra $\frak{h}$
must vanish
\begin{equation}
\label{coset.constr.bracket2}
\widetilde\eta [ \widetilde B_+ , \widetilde B_- ]_{\cal R} |_{\frak{h}} = \widetilde\eta ( [{\cal R}\widetilde B_+ , \widetilde B_-]+   [\widetilde B_+ , {\cal R} \widetilde B_-]|)  |_{\frak{h}}   = 0   \ .
\end{equation}
For $\tilde\eta\neq0$, this constraint is exactly in agreement with that found for the two-parameter  theories constructed in section \ref{lapp} to develop  a gauge invariance that reduces their dynamics to the coset.  When ${\cal R}$ entering in to eq.~\eqref{eq:symcosetdata} is identified with the projection of an ${\cal R}$-matrix on the group then this constraint is quite stringent; see discussion in section \ref{lapp}. 

From this rewriting one can see that the equations of motion can be written in terms of a classical Lax connection
\be
\boxed{
\label{LaxYB.coset}
{\cal L}_\pm=A_\pm+\mu^{\pm1}\left(\frac{\mathbb{1}}{\sqrt{\l_0}}\pm\tilde\eta\l_0^{\pm1/2}\,\R\right)\left(\mathbb{1}\pm\tilde\eta\R\right)^{-1}B_\pm\,,\quad
\mu\in\mathbb{C}\ .
}
\ee
Thus, there is a two-parameter family of integrable deformations labelled by $\tilde\eta$ and $\lambda_0$  for an arbitrary symmetric coset.

For comparison we may note that the $\tilde\eta=0$ limit returns to the known isotropic $\lambda$-deformation of a symmetric coset for which the Lax connection was given in  \cite{Hollowood:2014rla} as
\be
{\cal L}_\pm=A_\pm+\frac{\mu^{\pm1}}{\sqrt{\l_0}}\,B_\pm,\quad \mu\in\mathbb{C}\  .
\ee

\section{The $SU(2)$ paradigm}
\label{sec:SU2groupexamples}

In this section we examine the connection between the bi-YB $\eta$-deformations and the generalised $\lambda$-deformations considered in the preceding section.
In \cite{Klimcik:2002zj,Vicedo:2015pna,Hoare:2015gda} a PL T-duality transformation
followed by an analytic continuation related the single parameter $\eta$- and $\lambda$-deformations.
We expect that this also will be the case for multi-parameter deformations.
We explicitly demonstrate this in specific examples based on $SU(2)$ and $SU(2)/U(1)$.

For $\eta$-deformations based on $SU(2)$, the relevant Drinfeld double for performing PL T-duality is  $SU(2)\oplus E_3$.
In the appendix \ref{sec:AppDrinfeld} we provide explicit details for the parameterisations of the group elements, for the matrix realisation of generators, for the left-invariant one-forms
and for the group theoretic matrices $\Pi$ and $\tilde{\Pi}$ that enter into the definitions of the PL T-dual pairs in \eqref{eq:PLTdualpairs}.

 \subsection{The bi-YB on $SU(2)$}

The $SU(2)$ bi-YB $\sigma$-model  was shown in \cite{Hoare:2014pna} to be the Fateev model \cite{Fateev:1996ea}
and its RG flows can be read from \eqref{RG.biYB} for $c_G=4$ and $c^2=-1$.
This can be shown to be in agreement with the result for  the RG flow in \cite{Fateev:1996ea}.\footnote{The map between the parameters $(\eta,\zeta,t)$
and those of Fateev $(r, \ell,u)$ (defined after eq. (76) of  \cite{Fateev:1996ea})  is given by  \cite{Hoare:2014pna}
but needs to be slightly amended to include an overall tension $t$ needed for the renormalisability of the model:
\begin{equation*}
\eta^2=\frac{r}{u}(\ell u^{-1} +1) \ , \quad
\zeta^2=\frac{\ell}{u} (r u^{-1} +1) \ ,\quad
t=u\ .
 \end{equation*}  }

The target space geometry and anti-symmetric tensor of the bi-YB $\sigma$-model are given by \cite{Sfetsos:1999zm}
\begin{equation}
\begin{aligned}
   \mathrm{d}s^2 &=  \frac{1}{ t \Lambda} \left(  \mathrm{d}\psi^2+\mathrm{d}\theta^2 +\mathrm{d}\varphi^2+
2\cos\theta\mathrm{d}\psi\mathrm{d}\varphi  \right. \\ & ~~~~~~~~~~~~~~~~~~ \qquad \left. + \left( (\eta + \zeta \cos\theta) \mathrm{d}\varphi +
  (\zeta + \eta \cos\theta) \mathrm{d}\psi \right)^2  \right) \ ,
\\
\Lambda &=  1+\zeta^2 +\eta^2 + 2 \zeta\,\eta \cos\theta \ , \quad
H_3 = \mathrm{d}B_2 = 0 \ ,
\end{aligned}
\end{equation}
where the first line in $\mathrm{d}s^2$ corresponds to the round three sphere and we work in  the coordinates of appendix \ref{sec:AppDrinfeld}.

The bi-YB $\sigma$-model is symmetric under both left and right PL actions and here we perform a PL T-duality with respect to the left action.
This  results into a  dual $\sigma$-model whose target space geometry is
\begin{equation*}
\begin{aligned}
 {\mathrm{d}s^2} &= \frac{1}{\eta^2 t \Sigma} \left(-4 r \mathrm{d}r \mathrm{d}\chi  \left(\text{n}_+\left(r^2+1\right)-\text{m}_+ e^{2 \chi }\right)+4
  \mathrm{d}r^2 \left(\eta ^2 e^{2 \chi }+\text{n}_+r^2\right) \right. \\
   &\quad\quad  \left.
   +4 \eta ^2 r^2 e^{2 \chi }
\mathrm{d}\vartheta^2+\mathrm{d}\chi^2 \left(\Sigma -4 r^2 e^{2 \chi }\right) \right) \ , \\
 {B_2} &= \frac{4 r }{t \eta\S}\left(\mathrm{d} r \wedge \mathrm{d} \vartheta   \left(\text{n}_+\left(r^2+1\right)-\text{m}_+
   e^{2 \chi }\right)-2 r e^{2 \chi } \mathrm{d} \vartheta \wedge  \mathrm{d} \chi  \right)  \ , \\
{H_3} &= \frac{8 r e^{2 \chi} }{t \eta  \Sigma ^2}\left(\text{m}_- \Sigma +8 e^{2  \chi } \left(\eta ^2+\zeta ^2
   r^2\right)\right)  \mathrm{d}r \wedge \mathrm{d}   \chi \wedge \mathrm{d} \vartheta    \ ,\\
\Sigma &= \text{n}_- e^{4 \chi } -2 e^{2 \chi } \left(\text{m}_+-\text{m}_- r^2\right) +\text{n}_+
   \left(r^2+1\right)^2 \ ,
\end{aligned}
\end{equation*}
in which we work in the coordinates for dual model defined in the appendix \ref{sec:AppDrinfeld}. We have also introduced the constants
\begin{equation*}
\text{n}_\pm = 1 + (\zeta \pm \eta)^2 \ , \quad \text{m}_\pm = 1 \pm (\zeta^2 - \eta^2) \ .
\end{equation*}

We now perform an analytic continuation, which was used in the case of an $SU(2)/U(1)$ single $\eta$-deformation in \cite{Hoare:2015gda}
\begin{equation}\label{eq:cont}
r \mapsto i \sin \a \sin \beta \ , \quad e^\chi \mapsto \cos\a + i \sin\a \cos\b \ , \quad \eta \mapsto \frac{i (1-\l)}{(1+\l)} \ , \quad t
\mapsto \frac{ (1+\l)}{4 k(1-\l)}  \ .
\end{equation}
This results in the following expressions
\begin{equation}\label{eq:bilambdasu2}
\begin{aligned}
\frac{1}{k} {\mathrm{d}s}^2 & \mapsto \frac{1+\l}{1-\l} \left(1 + \zeta^2 \frac{(1+\l)^2 }{\Delta}  \sin^2\a \sin^2\b \right) \mathrm{d}\a^2   
+ \frac{1-\l^2}{\D} \sin^2\a  \mathrm{d}\Omega_2^2\\
  &+ 2\zeta  \frac{(1+\l)^2 }{\Delta}  \sin^2\a \sin \b\, \mathrm{d}\a\,\mathrm{d}\b   \ , \\
\frac{1}{k} {H}_3 &\mapsto  - {1\ov \D^2}  \left(  (4\l - \zeta^2(1+\lambda)^2)\Delta\right.\\
&\left.  +2  \left( (1-\l^2)^2 + \zeta^2 (1+\l)^4 \sin^2\a \sin^2\b \right)  \right)   \,
  \sin^2\alpha\,  \mathrm{d}\alpha \wedge \mathrm{vol}(S^2)\,, \\
   \Delta  &=  1+ \l^2 - 2\l \cos2\a + \zeta (\l^2-1)  \sin 2\a \cos\beta + \zeta^2 (1+\l)^2 \sin^2\a \cos^2\b\, ,
          \end{aligned}
\end{equation}
 with $\mathrm{d}\Omega_2^2  = \mathrm{d} \beta^2 + \sin^2\beta\, \mathrm{d}\vartheta^2$ and $\mathrm{vol}(S^2)=
 \sin\beta\, \mathrm{d} \beta \wedge  \mathrm{d}\vartheta$.
We note that the field strength $H_3$ is real, but the gauge potential produced by this continuation  includes an imaginary piece.

These $\s$-model background fields can be obtained via a generalised $\lambda$-deformation of the form,
\be\label{eq:lambdamatrix}
\lambda_{ab} = \left(
\begin{array}{ccc}
 \frac{\zeta ^2 (\lambda +1)^2+4 \lambda }{\zeta ^2 (\lambda +1)^2+4} & \frac{2
   \zeta  \left(1-\lambda ^2\right)}{\zeta ^2 (\lambda +1)^2+4} & 0 \\
- \frac{2 \zeta  \left(1-\lambda ^2\right)}{\zeta ^2 (\lambda +1)^2+4} &
   \frac{\zeta ^2 (\lambda +1)^2+4 \lambda }{\zeta ^2 (\lambda +1)^2+4} & 0 \\
 0 & 0 & \lambda  \\
\end{array}
\right)\ ,
\ee
with the group element entering into eq.~\eqref{tdulalmorev2}  parametrised as
\be
g = \left( \begin{array}{cc} \cos\a + i \sin \a \cos\b  &  \sin \a \sin \b e^{-i \vartheta} \\ - \sin \a \sin\b e^{i\vartheta} & \cos \a - i \sin\a \cos\b  \end{array}\right)  \ .
\ee
In this parametrisation one finds that in the $\zeta\rightarrow 0$ limit the known expressions for the $\lambda$-deformation  of $SU(2)$, see e.g. \cite{Sfetsos:2013wia,Sfetsos:2014cea}, are recovered.

The procedure of integrating out gauge fields in the derivation of the $\lambda$-deformation means that, when performed in a path integral, one should also complement the background fields with the dilaton factor
\be\label{eq:dilbilambdasu2}
e^{-2 \tilde{\Phi} -2 \Phi_0} = \Delta \ ,
\ee
in which $\Phi_0$ is simply a constant additive contribution.
One can verify that although these background fields do not solve the three-dimensional bosonic truncation of the supergravity equations, the dilaton beta function drastically simplifies to
\begin{equation*}
\begin{aligned}
\beta^{\widehat\Phi} &= \widehat R+ 4 \nabla^2 \widehat\Phi -4 (\nabla\widehat \Phi)^2 - \frac{1}{12} \widehat H^2\\
 &=   \frac{1}{2k (1-\l^2)(1+\l)^2 } \left( 8(1+2\l +2\l^3 +\l^4)+ 8\zeta^2 \l (1+\l)^2 - \zeta^4 (1+\l)^4 \right) \ .
\end{aligned}
\end{equation*}
 That such cancellations occur gives a strong hint that it may be possible to embed this two-parameter $\lambda$-deformation as a solution of supergravity along the lines of
\cite{Sfetsos:2014cea,Demulder:2015lva}. In these works the contribution to the dilaton beta-function is cancelled off against an opposite contribution that arises from performing the $\lambda$-deformation to a non-compact $SL(2, \mathbb{R})$. It seems likely that such a solution can be embedded into ten-dimensional IIB supergravity by including a spectator CFT on a $T^{4}$ and generalising the symmetry considerations leading to the RR-sector of \cite{Sfetsos:2014cea}.

We close this section with a rather appealing observation; the background fields of \eqref{eq:bilambdasu2} which was obtain from the bi-YB deformation by PL T-duality plus analytic contribution can be thought of as the $\lambda$-deformation of an $\eta$-deformed $\sigma$-model as described in section \ref{sec:lambdaonYB}.
To be precise, making use of the definition
\be
\quad  (\l^{-1})_{ab}= k^{-1}(\hat{E}+k\,\mathbb{1} )_{ab} \  ,
\ee
one finds that the PCM coupling matrix $\hat{E}$ corresponding to the $\lambda$-matrix \eqref{eq:lambdamatrix} is of the
YB $\sigma$-model form
\be
\hat E =  \frac{1}{\tilde t} ( \mathbb{1} - \tilde{\eta} {\cal R} )^{-1}  \ ,
\ee
with
\be
 \lambda = \frac{k\tilde t}{k\tilde t+1} \,,\quad  \tilde{\eta}  = -\zeta \frac{2k\tilde t + 1}{2k\tilde t } \
\ee
and where the $\R$-matrix of $SU(2)$ is given by
\be\label{eq:RSU(2)}
{\cal R} = \left( \begin{array}{ccc} 0 & 1 & 0 \\ -1 & 0 & 0 \\ 0 & 0 & 0  \end{array} \right) \ ,
\ee
in a basis where the generators are the normalised Pauli matrices $T_i = \frac{1}{\sqrt{2}} \sigma^i$.
This YB $\sigma$-model is renormalisable at one-loop in $\nicefrac1k$ with RG equations
\be
\dot\lambda=-2\frac{(1+ \tilde{\eta}^2)\lambda^2(1+ \tilde{\eta}^2\lambda^2)}{k(1+\lambda)^2}\,,\quad
\frac{ \tilde{\eta}\lambda}{1-\lambda}\quad{\rm and}\quad k\quad~{\rm constants}\,.
\ee
Study of the above RG-flow equations reveals that there is an arbitrary finite energy scale for which $\lambda\to0$. However,
the matrix \eqref{eq:lambdamatrix} does not tend to zero, so the conformal point is never reached.

In total we explicitly showed for the $SU(2)$ case, that the bi-YB and $\l$-deformed YB $\s$-models are related
with PL T-duality and analytic continuation with the parameters identified as follows:
\be
(t,\eta,\zeta)\mapsto(k,\tilde t,\tilde\eta):\quad
k=\frac{i}{4 t \eta}\ , \quad  \frac{k\tilde t}{k\tilde t+1}=\frac{i-\eta}{i+\eta} \ ,\quad  \tilde{\eta}  = -\zeta \frac{2k\tilde t + 1}{2k\tilde t } \,.
\ee

 \subsection{The bi-YB on $SU(2)/U(1)$}
\label{sec:SU2cosetexamples}
We now turn to an example based on the symmetric space $SU(2)/U(1)$.     The metric of the bi-YB $(\eta,\zeta)$-deformed $\sigma$-model given in \eqref{eq:biYBsigmamodelCoset}  reads
 \be
\label{biYBcoset}
\mathrm{d}s^2=\frac{1}{  t}\,
\frac{\mathrm{d}z\,\mathrm{d}\bar z}{1+(\zeta+\eta)^2+2z\bar z (1+\zeta^2-\eta^2)+z^2\bar z^2(1+(\zeta-\eta)^2)}\, .
\ee
Here we have adopted the parametrisation of $S^{2}$ used in section 4 of  \cite{Delduc:2013fga} and indeed for $\zeta=0$ this coincides with the  $\eta$-deformed $S^2$ i.e. eq.(4.2) of \cite{Delduc:2013fga}. The $\sigma$-model given by eq.\eqref{biYBcoset} is  one-loop renormalisable and the corresponding RG flow equations, given in general in   \cite{honer,Friedan:1980jf,Curtright:1984dz}, read
\be
\label{RG.biYB.coset}
\dot\eta=2t\,\eta(1-\zeta^2+\eta^2)\,,\quad \dot t=-2t^2(1-\zeta^2+\eta^2)\,, \quad \dot\zeta=-2\zeta\,t(1+\zeta^2-\eta^2)\,.
\ee
There are two invariants under the RG flow
\be
\label{jssgj}
t\,\eta\quad{\rm and}\quad \frac{1+\zeta^2+\eta^2}{\zeta\,\eta}={\rm constants}.
\ee

The metric of eq.~\eqref{biYBcoset} has also appeared in the studies of PL T-dual coset models associated with $S^2$
\cite{Klimcik:1996np,Sfetsos:1999zm}.  Defining $ a=1/\eta$ and $b=\zeta/\eta$, and  changing to stereographic coordinates $z=\cot(\theta/2)$ one finds  eq.~\eqref{biYBcoset} results in exactly the metric of eq.(3.16) of \cite{Sfetsos:1999zm} multiplied by an overall tension $T= \frac{1}{4  t\eta}$. Indeed, this system of RG flows along with its invariants were also in found \cite{Sfetsos:1999zm}, as a consistent truncation of those
 for the PL T-dual $\s$-model on the Drinfeld double $SU(2)\oplus E_3$.

We may now perform a PL T-dualisation of eq.~\eqref{biYBcoset} using \cite{Klimcik:1996np,Sfetsos:1999zm} resulting in a dual metric  given in eq. 3.18 of  \cite{Sfetsos:1999zm} multiplied by the overall tension $T$.   Performing a field redefinition\footnote{We change the variables of eq. 3.18 of  \cite{Sfetsos:1999zm} according to   \begin{equation}
z = \frac12( a+(b-1)^2a^{-1})\big( (p+q)^2-1\big) \ ,
\quad \rho = ( a+(b-1)^2 a^{-1})\sqrt{p^2-q^2 -1}\ . \nonumber
\end{equation}
  }
 and analytic continuation on coordinates and parameters
\be
i T= k \Longrightarrow t\,\eta=\frac{i}{4k}\,,\quad \frac{1}{\eta}=-i(1+2\alpha^2)\,,\quad \frac{\zeta}{\eta}=i\beta\,,\quad q\mapsto iq\, .
\ee
results in the $\sigma$-model  action
\be
\label{PLTWickbiYBcoset.action}
S=\frac{k}{\pi}\,\int  \mathrm{d}^2\s \,\frac{\left((1+2\alpha^2)^2+\beta^2\right) \del_+p\del_-p+
\beta\, \left(\del_+p\del_-q+\del_-p\del_+q\right)+\del_+q\del_-q}{(1+2\alpha^2)(1-p^2-q^2)}\, .
\ee
We now clarify an interesting point\footnote{Which we thank Ben Hoare for raising.}.  Prior to the PL T-dualisation, it turns out one can   effectively set the parameter $\zeta=0$ in eq.~\eqref{biYBcoset} by a transformation that  does not affect the global properties of the metric but simply rescales the
 overall coupling $t$.   One might at first think that the appearance of $\zeta$ is redundant, however, we shall now interpret the theory eq. \eqref{PLTWickbiYBcoset.action}  as $\lambda$-deformation in which the parameter $k$ is an integer quantised variable.   Rescaling $t$ so as to remove $\zeta$ would correspond to rescaling $k$ by an arbitary real number to remove $\beta$.   The deformation parameter $\beta  $ is thus significant and cannot be absorbed into a rescaling of $k$ without spoiling the topological nature of the overall coupling.

 This action can also be obtained as a generalised $\lambda$-deformation applied to an $\eta$-deformed $\sigma$-model.
 To see this let us begin by considering the PCM \eqref{eq:PCM2} equipped with the matrix
\be
\label{coset.PCM}
\hat E=\left(   \begin{matrix} \kappa^2 & \g & 0\\
          -\g & \kappa^2 & 0\\
0 & 0 & s^2 \end{matrix}   \right)\, , \quad \a^2=\frac{k\,\k^2}{\kappa^4+\g^2}\,,\quad \beta=-\frac{2k\g}{\kappa^4+\g^2} \ .
\ee
As in \cite{Sfetsos:2013wia} in order to recover a two-dimensional model we take the limit $s^2 \rightarrow 0$.
 This will implement the truncation of the $SU(2)$ PCM to just the $SU(2)/U(1)$ coset.  Explicitly, if one parametrises the $SU(2)$  group element as
\be
g=e^{i(\varphi_1-\varphi_2)\s_3/2}e^{i\omega\s_1}e^{i(\varphi_1+\varphi_2)\s_3/2}\, ,
\ee
then in the generalised $\lambda$-deformed theory \eqref{tdulalmorev2}, where $\lambda^{-1}=k^{-1}(\hat E+k)$, one finds that after taking the limit $s\to0$ the coordinate $\varphi_2$ drops out of the action altogether and can be fixed to any value we choose. This reflects a residual $U(1)$  gauge invariance remaining after fixing the group element of the PCM, $\hat{g} = \mathbb{1}$. The resulting $\lambda$-deformed theory matches exactly the one in \eqref{PLTWickbiYBcoset.action} upon changing to algebraic coordinates
\be
\label{transf}
p=\cos\omega\cos\varphi_1\,,\quad q=\cos\omega\sin\varphi_1 \  .
\ee
This $\lambda$-deformed action has an interpretation in terms of the $SU(2)/U(1)$ CFT deformed by a para-fermionic bilinear generalising the results of \cite{Sfetsos:2013wia}.
For small $\a$ and $\b$ the dominant term in \eqref{PLTWickbiYBcoset.action} corresponds to the exact $SU(2)/U(1)$ CFT
\be
\label{CFTS2}
S_{\rm CFT}=\frac{k}{\pi}\int  \mathrm{d}^2\s\, \left(\del_+\omega\del_-\omega+\cot^2\omega\, \del_+\varphi_1\del_-\varphi_1\right)\ ,
\ee
where we have performed the change of variables \eqn{transf}.
The full action \eqref{PLTWickbiYBcoset.action}  can be expressed in terms of this CFT and bilinears of para-fermionic operators defined by
\be
\begin{aligned}
\psi &: =  e^{-i \tilde \varphi_1 }\frac{\partial_+(p-iq) }{\sqrt{1-p^2-q^2}} =
\big(\del_+ \om + i \cot\om\
\del_+ \varphi_1\big) e^{-i(\varphi_1+\tilde \varphi_1)} \ ,    \\
\bar \psi &: = e^{i \tilde \varphi_1 }\frac{\partial_-(p-iq) }{\sqrt{1-p^2-q^2}} = \big( \del_- \om + i
\cot\om\   \del_- \varphi_1\big) e^{-i(\varphi_1-\tilde\varphi_1)} \ ,
\end{aligned}
\ee
and their complex conjugates $\psi^\dag$ and $\bar\psi^\dag$.  Here $\tilde\varphi_1$ is a non-local function of $\omega$ and $\varphi_1$, that dresses the operators to ensure conservation $ \del_- \psi= \del_+ \bar\psi =0$.  With these  the action \eqref{PLTWickbiYBcoset.action} can be expressed as
\be
\label{lambda.coset.action}
\begin{split}
&S=c_1\,S_{\rm CFT}+\frac{k c_2}{\pi}\int  \mathrm{d}^2\s\, \left(\psi\bar \psi + \psi^\dag\bar \psi^\dag\right)+
\frac{k c_3}{\pi}\int  \mathrm{d}^2\s\,  \left(\psi\bar \psi - \psi^\dag\bar \psi^\dag\right)\,,\\
&c_1=1+\frac{4\a^4+\b^2}{2(1+2\a^2)}\,,\quad
c_2=\frac{4\a^2(1+\a^2)+\b^2}{4(1+2\a^2)}\,, \quad
c_3=\frac{i \beta}{2(1+2\a^2)}\,.
\end{split}
\ee
In an expansion of small $\a$ and $\b$ one sees that  eq.~\eqref{lambda.coset.action}  perturbs the exact CFT action \eqref{CFTS2}   by para-fermionic bilinears which act as relevant operators since the para-fermions have conformal dimension $1-1/k$.
In that respect the $\s$-model \eqn{lambda.coset.action} is renormalisable at one-loop in $1/k$ with
\be
\dot\lambda_0=-\frac{\l_0}{k}\,,\quad \dot\g=\frac{\g}{k}\,,\quad  k={\rm constant}\,,\quad \l_0=\frac{k}{k+\kappa^2}\,.
\ee
These RG equations imply in the UV that the  parameters $\a,\b$, defined in \eqref{coset.PCM}, go to zero and the model flows to the WZW model.

\no
Towards the IR the parameter $\l_0$ tends to unity and one has to perform a stretching of the coordinates in order to
make sense of the geometry. This limit is also achieved by letting in \eqref{lambda.coset.action} the following rescaling
\be
\varphi_1=\frac{x_1-\g}{2k}\,,\quad \omega=\frac{x_2}{2k}\,,
\ee
followed by the limit $k\to\infty$. As we can see the running of $\g$ is irrelevant since it can be absorbed into a field redefinition
of $\varphi_1$. The end result of this limiting procedure is
\be
\mathrm{d}s^2=\frac{1}{2}\left(\kappa^2\,\frac{\mathrm{d}x_1^2}{x_2^2}+
\frac{1}{\kappa^2}\left(\mathrm{d}x_2+\frac{x_1}{x_2}\mathrm{d}x_1\right)^2\right)\ ,
\ee
which is the non-Abelian T-dual of the PCM on $S^2$ as expected on general grounds \cite{Sfetsos:2013wia}.

\subsubsection*{Integrability}

Since this $\lambda$-deformed theory falls in the class considered in sec.~\ref{sec:symcoset} we have already proven its integrability.  However, we present in appendix \ref{ksks} an explicit demonstration of its
integrality in the hope that the reader may find, as we did, it to be illuminating.

\section*{Acknowledgements}

We  thank the University of Bern for hospitality in hosting the workshop "Integrable $\eta$- and $\lambda$-deformations in supergravity"  where this work was completed and first presented.  We thank the participants of this workshop for raising many interesting comments as well as Ben Hoare, Luis Miramontes and Beno\^it Vicedo 
for useful communications after a first version of the paper appeared. The research of K. Sfetsos is implemented
under the \textsl{ARISTEIA} action (D.654 of GGET) of the \textsl{operational
programme education and lifelong learning} and is co-funded by the
European Social Fund (ESF) and National Resources (2007-2013).
The research of K. Siampos has been supported by the Swiss National Science Foundation.
K.~Siampos also acknowledges the \textsl{Germaine de Stael}
France--Swiss bilateral program (project no 32753SG) for financial support.
The work of D.C. Thompson was supported
in part by FWO-Vlaanderen through project G020714N and postdoctoral mandate
12D1215N, by the Belgian Federal Science Policy Office through the Interuniversity
Attraction Pole P7/37, and by the Vrije Universiteit Brussel through the Strategic
Research Program "High-Energy Physics".

\appendix

\section{Properties of the $\R$ matrix}

\label{Rappendix}

This appendix will be devoted to a brief summary of the modified classical YB equation
and properties of the $\R$ matrix.

Consider a semisimple Lie group $G$, a Lie algebra $\frak{g}$, and a
matrix ${\cal R}$ (an endo-morphism of $\frak{g}$),  assumed to be anti-symmetric  with respect to the Killing form on $\frak{g}$, which defines a bracket
\be
\label{Rbracket}
[A,B]_{\R}=[\R A,B]+[A,\R B] \ ,  \quad \forall  A,B \in \frak{g} \, .
\ee
A sufficient condition for \eqref{Rbracket} to satisfy  the Jacobi identity is
the modified classical YB equation (mYB)
\be
[\R A, \R B] - \R[A,B]_\R = -c^2[A, B]  \ ,  \quad \forall  A,B \in \frak{g} \,,\quad c\in\mathbb{C}\,.
\ee
Note that $\R$ matrix can be rescaled and this results to three different distinct classes:
$c=0$, $c=1$ and $c=i$.

Expanding in an arbitrary basis we can write
\be
A= A_a T_a\ ,\quad \R A = (\R A)_a T_a = \R_{ab} A_b T_a \ ,
\ee
and using that $\R_{ab}=-\R_{ba}$, we find explicitly

\be
\label{mYB1}
c^2\,f_{abc} + \R_{ad} \R_{be} f_{dec} +\R_{bd} \R_{ce} f_{dea} + \R_{cd} \R_{ae} f_{deb}=0\,,
\ee
or
\be
\label{mYB2}
\R_{ad}\tilde f_{bcd}+\R_{cd}\tilde f_{abd}+\R_{bd}\tilde f_{cad}=2c^2\,f_{abc}\,,\quad
 \tilde f_{abc}=\R_{ad}f_{bdc}-\R_{bd}f_{adc}=-\tilde f_{bac}\,.
\ee
The $f_{abc}, \tilde f_{abc}$ are the structure constants of the usual and
the $\R$-bracket \eqref{Rbracket} respectively, satisfying the Jacobi identities
\be
\label{Jacobi}
 f_{abd}  f_{dce}+ f_{cad}  f_{dbe}+ f_{bcd}  f_{dae}=0\,,\quad
 \tilde f_{abd} \tilde f_{dce}+\tilde f_{cad} \tilde f_{dbe}+\tilde f_{bcd} \tilde f_{dae}=0\,,
\ee
and identically satisfying the relation
\be
\label{mixedJacobi}
f_{abd}\tilde f_{ced}+f_{dac}\tilde f_{deb}-f_{dbc}\tilde f_{dea}-f_{dae} \tilde f_{dcb}+f_{dbe} \tilde f_{dca}=0\,.
\ee
The choice of matrix $\R$ in fact specifies a Drinfeld double
 \begin{equation}
 \frak{d} = \frak{g} \oplus \frak{g}_R\,,
 \end{equation}
as $f_{abc},\tilde f_{abc}$ satisfy their Jacobi identities \eqref{Jacobi} and the mixed one \eqref{mixedJacobi}.

In what follows, we shall focus on $c=i$, referred to as the complex branch or the "non-split case".  In this case the Drinfeld double is just the complexification  $\frak{d}  = \frak{g} \oplus \frak{g}_R=\frak{g}^\mathbb{C}$  viewed as a Lie algebra.  On $\frak{g}^\mathbb{C}$ we have an inner product
 \[
 \langle A+ i B , A'   + i B' \rangle = \Im (A+i B, A'+ i B')  \ ,
 \]
 with respect to which $\frak{g}$ is a maximal isotropic {\em and} when $\R$ is anti-symmetric w.r.t. $( \cdot, \cdot)$ so is $\frak{g}_R$.
This Drinfeld double admits an Iwasawa decomposition
\be
D= G^\mathbb{C}= G A N=A N G\,,
 \ee
 where an element  $A N$ can be expressed in terms of positive roots and a Hermitian element
 \[
 \mathrm{e}^\varphi \exp \sum_{\alpha>0 } v_\alpha E_\alpha\,.
 \]
For $D=SL(n, \mathbb{C})$ groups, $AN$ can be identified with
the group of upper triangular matrices of determinant 1 and with positive
numbers on the diagonal and $G = SU (n)$.

\section{The $\frak{su}(2) \oplus \frak{e}_3$ Drinfeld Double}

We follow with small modifications the parametrisation of \cite{Sfetsos:1999zm} and rederive the necessary for
our purposes results.

\label{sec:AppDrinfeld}
We use a block diagonal matrix representation for generators of  $\frak{su}(2)$ and  $\frak{e}_3$ given respectively by
 \begin{equation*}
 \begin{split}
&T_1= \frac{1}{2}(\s^1 , \s^1) \ , \quad T_2 = \frac{1}{2}(\s^2 , \s^2) \ , \quad T_3=  \frac{1}{2}(\s^3 , \s^3) \ ,  \\
&\tilde{T}^1 = i ( \s^+ , -\s^-) \ , \quad \tilde T^2 = (\sigma^+  , \sigma^-)  \ , \quad \tilde{T}^3 = \frac{i}{2} ( \s_3 , - \s_3) \ ,
\end{split}
 \end{equation*}
 where $\s^\pm = \frac{1}{2} (\s^1 \pm i \s^2)$.   We define an inner product on $\frak{su}(2) \oplus \frak{e}_3$ by
  \begin{equation*}
 \langle X , Y \rangle  = -i\,  \mathrm{tr} \left(P_u X P_u Y - P_d X P_d Y \right) \ ,
  \end{equation*}
 where $P_u$ projects onto the top left two by two block and $P_d$ onto the bottom right.  If we let $T^{A} = \{ T_{i} , \tilde{T}^{j} \}$ with $A = 1\dots 6$, $i,j= 1,2,3$, be a basis for the generators of the double then
 \begin{equation*}
 \langle T_{A}, T_{B} \rangle =   \left( \begin{array}{cc} 0 & \mathbb{1}_{3}  \\ \mathbb{1}_{3} & 0 \end{array} \right) \ ,
 \end{equation*}
  indicating that $\frak{su}(2)$ and $\frak{e}_{3}$ span mutually orthogonal maximal isotropic subspaces with respect to this inner product.

    We parametrise a group element as
 \begin{equation*}
 \begin{aligned}
 & g_0 = \exp(i/2 \varphi \s^3) \exp(i/2 \theta \s^2) \exp(i/2 \psi \s^3) \ , \quad  g_{SU(2)} =  (g  ,g ) \ ,  \\
  & g_{+} = \left(\begin{array}{cc} \mathrm{e}^{\chi/2}  & \mathrm{e}^{-\chi/2}(y_1 - i y_2) \\ 0 &  \mathrm{e}^{-  \chi/2}    \end{array}\right) \ , \quad
    g_{-} = \left(\begin{array}{cc} \mathrm{e}^{-\chi/2}  & 0  \\  - \mathrm{e}^{-\chi/2}(y_1 +  i y_2)  &   \mathrm{e}^{   \chi/2}    \end{array}\right)
     \ , \quad g_{E_3} = ( g_+, g_-)\,.
    \end{aligned}
 \end{equation*}

Using this parameterisation one finds that the left-invariant one-forms for $\frak{su}(2)$ defined by 
$L^{i} = - i \langle g_{SU(2)}^{-1} \mathrm{d} g_{SU(2)} ,  \tilde{T}^{i}\rangle$ are given by
 \begin{equation*}
 \begin{aligned}
& L^1 =  \sin \theta  \cos \psi   \mathrm{d} \varphi  -  \sin \psi  \mathrm{d} \theta   \ , \quad
 L^2 =\sin \theta  \sin \psi
\mathrm{d} \varphi  + \cos \psi    \mathrm{d} \theta\ ,  \quad
  L^3 =  \cos \theta  \mathrm{d} \varphi  + \mathrm{d} \psi \ ,
    \end{aligned}
 \end{equation*}
whilst those of  $\frak{e}_{3}$ defined by $\tilde L_{i} = - i \langle g_{E_3}^{-1} \mathrm{d} g_{E_3},  T_{i}\rangle$ are
 \begin{equation*}
 \begin{aligned}
  & \tilde{L}_1 = -e^{-\chi }  \mathrm{d} y_1\ , \quad   \tilde{L}_2 = -e^{-\chi }  \mathrm{d} y_2 \ ,
\quad     \tilde{L}_3 =  - \mathrm{d} \chi\,.
    \end{aligned}
 \end{equation*}
The group theoretic matrices defined in \eqref{eq:pidef} are
\begin{equation*}
\begin{split}
&\Pi = \left(
\begin{array}{ccc}
 0 & -2 \sin ^2\frac{\theta }{2} & -\sin \theta  \sin \psi  \\
 2 \sin ^2\frac{\theta }{2} & 0 & \cos \psi  \sin \theta  \\
 \sin \theta  \sin \psi  & -\cos \psi  \sin \theta  & 0 \\
\end{array}
\right) \ , \\
\tilde{\Pi} &=
\left(
\begin{array}{ccc}
 0 & \frac{1}{2} e^{-2 \chi } \left(-y_1^2+e^{2 \chi }-y_2^2-1\right) &
   -e^{-\chi } y_2 \\
 \frac{1}{2} e^{-2 \chi } \left(y_1^2-e^{2 \chi }+y_2^2+1\right) & 0 &
   e^{-\chi } y_1 \\
 e^{-\chi } y_2 & -e^{-\chi }y_1 & 0 \\
\end{array}
\right) \ .
\end{split}
\end{equation*}
We will further define $y_1 + i y_2  = r e^{i \vartheta}$.

\section{Integrability of the generalised $\l$-deformed $SU(2)/U(1)$}
\label{ksks}

The starting point of our proof are the equations of motion for $A_\pm$ \eqref{eom.A},
where $\lambda^{-1}$ is given in terms of \eqref{coset.PCM} for $s=0$
\be
\lambda^{-1}=k^{-1}(\hat E+k)=\left(   \begin{matrix} \l_1 & \l_2 & 0
\\
          -\l_2 & \l_1 & 0\\
0 & 0 & 1 \end{matrix}   \right)\,,\quad \l_1=\frac{k+\kappa^2}{k}\,,
\quad \l_2=\frac{\g}{k}\,.
\ee
Plugging the latter in \eqref{eom.A} and solving we find
\be
\begin{split}
&\del_\pm A_\mp^1=-A_\mp^2\,A_\pm^3\,,\quad \del_\pm A_\mp^2=A_\mp^1\,A_\pm^3\ ,
\\
&\del_+A_-^3-\del_-A_+^3=\l_1(A_+^1A_-^2-A_+^2A_-^1)-\l_2(A_+^1A_+^2+A_-^1A_-^2)\,.
\end{split}
\ee
Classical integrability is ensured by rewriting the
equations of motion in terms of a spectral dependent classical Lax pair
\be
\label{Lax}
{\cal L}_\pm=(c^1_\pm\,A_\pm^1+c^2_\pm\,A_\pm^2)\,T_1+
(d^1_\pm\,A_\pm^1+d^2_\pm\,A_\pm^2)\,T_2+ A_\pm^3\,T_3\,,
\ee
where the various coefficients are given by
\be
c_\pm^1=\sqrt{\l_1}\,\tilde\mu^{\pm1}\,,\quad c_+^2=-d_+^1=-\frac{\l_2}{\sqrt{\l_1}}\,\tilde\mu\,,\quad
c_-^2=d_-^1=0\,,\quad
d_\pm^2=\sqrt{\l_1}\,\tilde\mu^{\pm1}\,,
\ee
where  $\tilde\mu\in\mathbb{C}$.  This is in agreement with Lax pair presented in the general discussion \eqref{LaxYB.coset} when
$(\l_1,\l_2,\tilde\mu)\mapsto(\tilde\eta,\l_0,\mu)$:
\begin{equation}
\lambda_1=\frac{1+\tilde\eta^2\l_0}{\l_0(1+\tilde\eta^2)}\,,\quad \tilde\mu=\mu\sqrt{\frac{1+\tilde\eta^2\l_0}{1+\l_0}}\,,\quad
\l_2=\frac{\tilde\eta}{1+\tilde\eta^2}\frac{1-\l_0}{\l_0}\,,
\end{equation}
for the $\R$-matrix given by the projection into the coset of the $SU(2)$ $\R$-matrix i.e. the top left 2$\times$2 block of \eqref{eq:RSU(2)}.

Specialising for $\l_2=0$, we find the Lax pair for an isotropic deformation \cite{Hollowood:2014rla}
\be
{\cal L}_\pm=\sqrt{\l_1}\,\tilde\mu^{\pm1}\left(A_\pm^1T_1+A_\pm^2T_2\right)+A_\pm^3T_3\,.
\ee
 Since there is an equivalence at the level of equations of motion, this also shows the integrability of the model eq.~\eqref{eq:biYBsigmamodelCoset} for this specific case of $SU(2)/U(1)$.

\end{document}